\documentclass{elsart4-1}

\usepackage{graphics}
\usepackage{graphicx}
\usepackage{float}
\usepackage{floatflt}
\DeclareGraphicsExtensions{.jpg,.pdf,.png,.eps}
\usepackage{epsfig}

\usepackage{amssymb}

\usepackage[english,francais]{babel}

\newtheorem{e-proposition}[theorem]{Proposition}

\newtheorem{e-definition}[theorem]{Definition\rm}

\renewcommand{\theequation}{\arabic{equation}}
\def\og{\leavevmode\raise.3ex\hbox{$\scriptscriptstyle\langle\!\langle$~}}
\def\fg{\leavevmode\raise.3ex\hbox{~$\!\scriptscriptstyle\,\rangle\!\rangle$}}

\begin{document}

\centerline{Physics or Astrophysics/Header}
\begin{frontmatter}


\selectlanguage{english}
\title{Reduction of phonon mean free path: from low temperature physics to room temperature applications in thermoelectricity}


\selectlanguage{english}
\author[authorlabel1]{Olivier Bourgeois},
\ead{olivier.bourgeois@neel.cnrs.fr}
\author[authorlabel1]{Dimitri Tainoff},
\author[authorlabel1]{Adib Tavakoli},
\author[authorlabel1]{Yanqing Liu},
\author[authorlabel1]{Christophe Blanc},
\author[authorlabel2]{Mustapha Boukhari},
\author[authorlabel2]{Andr\'e Barski},
\author[authorlabel2]{Emmanuel Hadji}

\address[authorlabel1]{Institut N\'EEL, CNRS, 25 avenue des Martyrs, F-38042 Grenoble, France \\
Univ. Grenoble Alpes, Inst NEEL, F-38042 Grenoble, France}
\address[authorlabel2]{Institut Nanosciences et Cryog\'enie, SP2M, CEA-UJF, 17 Rue des Martyrs, Grenoble 38054, France}


\medskip
\begin{center}
{\small Received *****; accepted after revision +++++}
\end{center}

\begin{abstract}
It has been proposed for a long time now that the reduction of the thermal conductivity by reducing the phonon mean free path is one of the best way to improve the current performance of thermoelectrics. By measuring the thermal conductance and thermal conductivity of nanowires and thin films, we show different ways of increasing the phonon scattering from low temperature up to room temperature experiments. It is shown that playing with the geometry (constriction, periodic structures, nano-inclusions), from the ballistic to the diffusive limit, the phonon thermal transport can be severely altered in single crystalline semiconducting structures; the phonon mean free path being in consequence reduced. The diverse implications on thermoelectric properties will be eventually discussed.
\vskip 0.5\baselineskip

\selectlanguage{francais}
\noindent{\bf R\'esum\'e}
\vskip 0.5\baselineskip
\noindent
{\bf Reduction du libre parcours moyen des phonons: de la physique des basses temp\'eratures aux applications en thermo\'electricit\'es \`a l'ambiante. }
Il a \'et\'e propos\'e depuis longtemps maintenant que diminuer la conductivit\'e thermique en r\'eduisant le libre parcours moyen des phonons est une des meilleures facons d'am\'eliorer les performances actuelles des mat\'eriaux thermo\'electriques. En mesurant la conductance thermique et la conductivit\'e thermique de nanofils et de couches minces, nous montrons diff\'erentes mani\`eres d'augmenter la diffusion des phonons des basses temp\'eratures jusqu'\`a temp\'erature ambiante. Il est montr\'e qu'en jouant sur la g\'eom\'etrie (constriction, structures p\'eriodiques, nano-inclusions), \`a partir de la limite balistique jusqu'\`a la limite diffusive, le transport thermique phononique peut \^etre de facon significative alt\'er\'e dans des structures monocristallines semiconductrices; le libre parcours moyen des phonons \'etant r\'eduit en cons\'equence. Enfin, les diverses implications sur les propri\'et\'es thermo\'electriques seront discut\'ees.

 \vskip 0.5\baselineskip
\noindent{\small{\it Mots-cl\'es~:} phonon; transport thermique;thermo\'electricit\'e, semiconducteur, nanofil, inclusions }}
\end{abstract}
\end{frontmatter}

\selectlanguage{francais}
\section*{Version fran\c{c}aise abr\'eg\'ee}

\selectlanguage{english}
\section{Introduction}
\label{sec1}

In the two last decade, materials structured at the nanoscale having low thermal conductivity have attracted a lot of attention  due to their high potential thermoelectric applications \cite{Cahill2003,shakouriReview,Cahill2014,leb2014}. Indeed, as proposed by Dresselhaus \textit{et al.} in 1993, nanostructuration can significantly improve the thermoelectric efficiency \cite{dress1,dress2}. This efficiency of thermoelectric materials is given by the dimensionless figure of merit $ZT=S^2T\sigma/k$ where $S$ is the Seebeck coefficient, $\sigma$ the electrical conductivity and $k$ the thermal conductivity. If we focus only on the thermal properties, the $ZT$ will be increased by reducing the phonon thermal transport. This reflects the fact that the heat source and the cold source must be as thermally isolated as possible to prevent too much heat link that will degrade the overall performances of the thermoelectric modules.

\begin{figure}
\begin{center}
 \includegraphics[width=10cm]{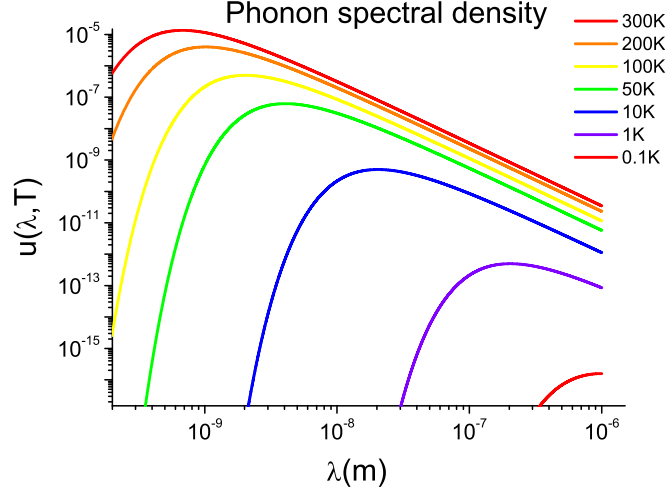}
 \end{center}
 \caption{Phonon spectral density as a function of the wavelength for different equilibrium temperature of a black body. The dominant phonon wavelength for each temperature corresponds to the maximum of the curves. Typically the dominant phonon wave length is of the order of 1~nm at 300~K and 100~nm around 1~K.} 
 \label{fig0}
\end{figure}

Since concerning electron, the thermal and the electronic properties are coupled through the Wiedemann-Franz law, the only way of reducing the thermal transport requires the reduction of the \textit{phonon} thermal conductivity. Two major directions are available to act on the phonon properties:  incoherent or coherent effects. On one hand, incoherent effects are ruled by the phonon mean free path (MFP) then the reduction is obtained by adding scattering centers, by reducing the size, having rough surface or by adding nano-inclusions. On the other hand coherent effects will manifest themselves through the reduction of the group velocity (band flattening), opening of band gaps or having destructive interferences. Two significant length scales govern the phonon physics involving incoherent or coherent effect: the phonon mean free path ($\Lambda$) and the dominant phonon wavelength given by:
\begin{equation}
\lambda_{dom}=\frac{h v_s}{2.82 k_B T}
\label{ldom}
\end{equation}
where $h$ the Planck constant, $k_B$ the Boltzmann constant and $v_s$ is the speed of sound. This length is indeed temperature dependent, increasing as the temperature decreases (see Fig.~\ref{fig0}).

Here, we will focus our attention on the experimental attempts to reduce of phonon mean free path in single crystal semiconducting materials by playing on the geometry of the surface or on the internal structure. Other means based on coherent effects will be discussed at the end of the paper. There is actually two complementary approaches to realize an efficient phonon engineering at the nanoscale: the top-down one based on e-beam lithography to make small structures like nanowire and membrane and the second one, the bottom-up, generally based on specific growth of nanostuctured materials (bulk or nanowire) using for instance molecular beam epitaxy (MBE).    

These two important limits will be experimentally addressed in this paper summarizing the recent findings of our group: the low temperature one set by boundary scattering limit where both important lengths  ($\Lambda$ and $\lambda_{dom}$) play a role and the room temperature limit where for the time being only the mean free path can be engineered efficiently.

\section{Towards the ballistic limit: thermal transport in Si nanowires}
\label{sec2}

In low dimensional systems, the surface scattering of phonons plays a crucial role in the understanding of thermal transport \cite{cas1939,ziman,berman}. The so-called Casimir regime, occurs when the bulk phonon MFP is much bigger than the actual size of the nanosystem under study. In that case, the phonon MFP is only set by the scattering on the surfaces. This limit is reached at low enough temperature when other scattering mechanisms do not dominate (phonon-phonon interaction, electron-phonon interaction, scattering on impurities etc...). As a result, effects linked to surface roughness have a significant impact on the phonon transport \cite{Heron1,Heron2}. These effects can be simple incoherent scattering when the phonon wavelength is much bigger than the mean asperity or, as the temperature is lowered, may become wavelength dependent. Indeed, if the wavelength is bigger than the mean roughness, specular reflections of phonons may occur on the surface. If elastic scattering is involved, the energy is conserved leading to an increase of the MFP.

\subsection{Normal phonon transport in Si nanowires at low temperature}

In order to understand the thermal transport at low temperature a summary of the Casimir and Ziman models is necessary. In the Casimir model, \cite{cas1939,berman}, the thermal conductance is given by:
\begin{equation}
K_{Cas}=3.2 \times 10^{3} \left( \frac{2\pi ^{2}k_{B}^{4}}{5 \hbar^{3} v_{s}^{3}} \right) ^{(2/3)} \frac{e\times w \Lambda_{Cas}}{L}T^{3}=\beta_{Cas} T^{3},
\label{Caseq}
\end{equation}
where $k_{B}$ is the Boltzmann constant, $\hbar$ the Planck constant and $v_{s}$ refers to the sound velocity. $\Lambda_{Cas}=1.12 \sqrt{e\times w}$ is the Casimir MFP of the phonons in a nanowire having a rectangular section, $e$ refers to the thickness and $w$ to the average width of the nanowire; $L$ being its length. This formula is obtained when the mean free path of phonons is only limited by the cross-section of the wire, i.e. by boundary scattering. Indeed, in this limit, each phonon hitting the rough surface is assumed to be absorbed an re-emitted in all direction and energy following the black body radiation law (see Fig.~\ref{fig0}). In other words, in this Casimir limit, the surface is considered as infinitely rough whatever the phonon wavelength.

\begin{figure}
\begin{center}
 \includegraphics[width=12cm]{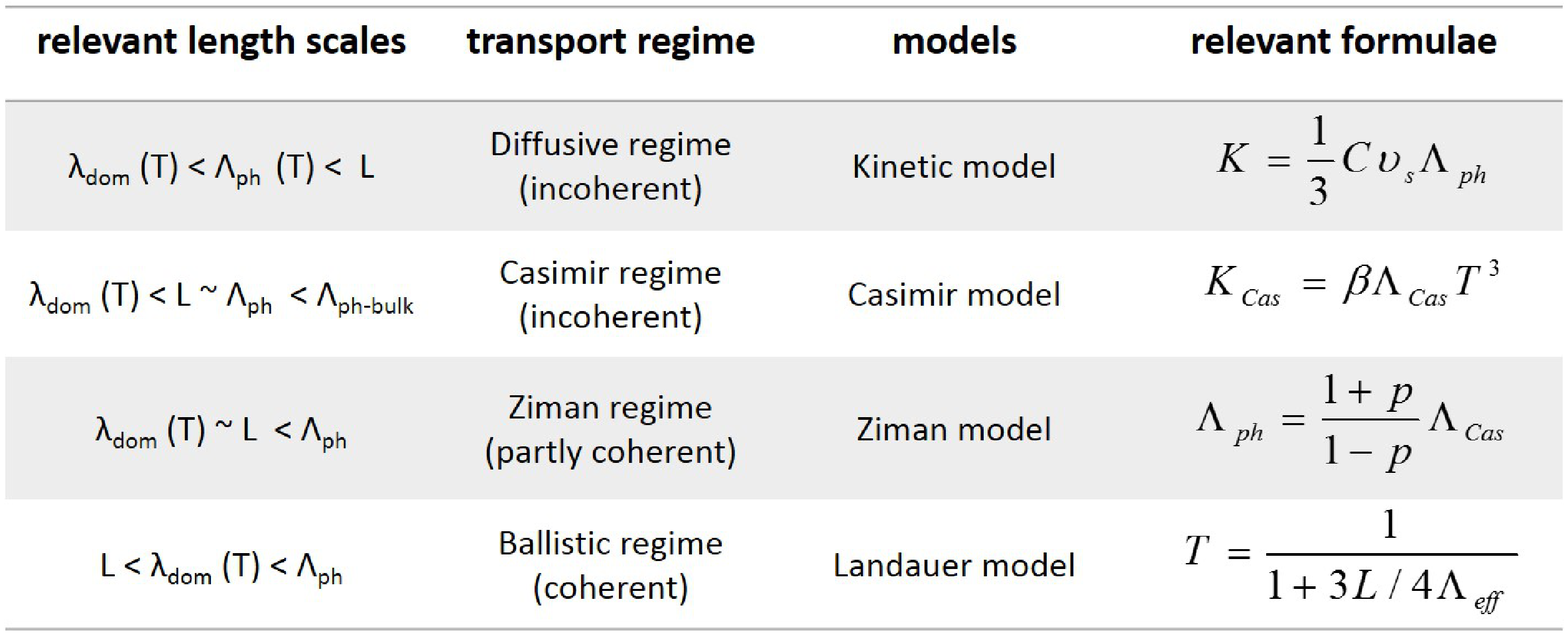}
 \end{center}
 \caption{Classification of the different regimes of phonon transport in nanoscaled systems at low enough temperature, typically below 30K for silicon. $L$ is the sample length, $\Lambda_{ph}$ is the inelastic phonon mean free path, and $\lambda_{dom}$ the dominant phonon wave length. Concerning the Landauer limit, the parameter $T$ denotes the transmission coefficient between a nanowire and its heat reservoir. Other important lengths are explained in the text.} 
 \label{table1}
\end{figure}

As the temperature is lowered, the dominant phonon wavelength actually grows (see equation~\ref{ldom}) and becomes eventually larger than the roughness amplitude; then specular reflections become more probable and cannot be neglected like in the Casimir model. To take this trend into account, the Casimir mean free path should be replaced in equation~\ref{Caseq} by an effective MFP $\Lambda_{eff}$ as proposed by Ziman and coll. \cite{ziman,berman}:
\begin{equation}
\Lambda_{eff}=\frac{1+p}{1-p} \Lambda_{Cas}
\label{zimfp}
\end{equation}
where $p$ is the parameter describing the probability for a phonon to be specularly reflected at the surfaces. The value of this parameter will depend on both the wavelength of the phonons and the roughness of the wire surface.
The Casimir purely diffusive case is given by $p=0$ and the purely specular one, when $p=1$, will be referred to as the ballistic limit; the MFP is diverging to infinite. The Ziman model takes into account a value for $p$ that is different from zero, as the temperature is lowered the phonon mean free path can be bigger than the section of the nanowire. The different phonon transport regimes and their associated models are summarized in the Table given in Fig.~\ref{table1}.

As a conclusion, the thermal transport in silicon nanowires having section of 100~nm will be well described by the Casimir and then at lower temperature by the Ziman model as it will be seen in the next section. Indeed, the length $L$ is smaller than the bulk phonon mean free path $\Lambda_{ph-bulk}$ in Si, and on the same order of magnitude as the dominant phonon wavelength. At even lower temperature, the fully ballistic limit should be described by a Landauer type model where the thermal transport is given by transmission coefficients between the nanowire and the heat bath. This goes clearly beyond our focus in this paper; there is currently no experimental data in this regime.

\begin{figure}
\begin{center}
 \includegraphics[width=14cm]{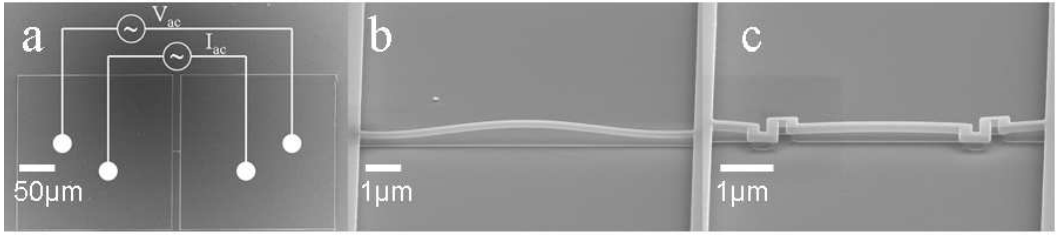}
 \end{center}
 \caption{a SEM images of: a) a suspended nanowire linked to the two pads where the electrical connections permit the 3$\omega$ measurements b) a straight nanowire c) corresponds to the side view of a S-shaped nanowire.} 
 \label{fig1}
\end{figure}

\begin{figure}
\begin{center}
 \includegraphics[width=10cm]{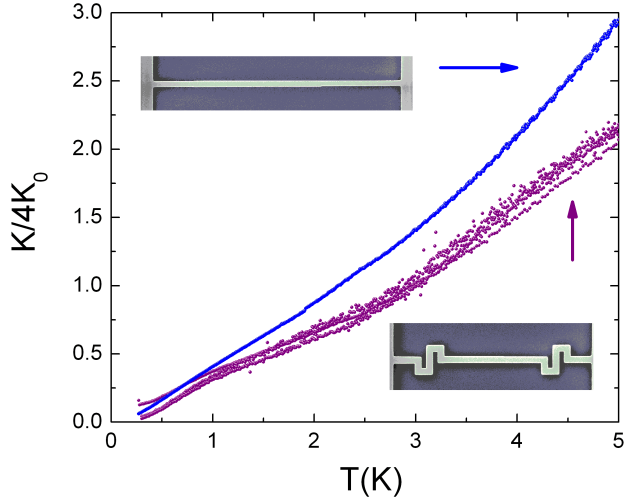}
 \end{center}
 \caption{Thermal conductance of 5$\mu$m long nanowires normalized to four times the universal value of thermal conductance versus temperature. The thermal conductance of the straight nanowire is much bigger than the thermal conductance of the S-shape nanowires. The presence of the serpentine section stops the transmission of ballistic phonons, reducing the overall phonon mean free path.} 
 \label{fig2}
\end{figure}

\subsection{Phonon blocking with S-shaped chicanes in Si nanowires}

We will illustrate in the following section these different scatterings and phonon transport regimes in three situations: firstly the thermal transport in a normal straight silicon nanowire, secondly in a Si nanowire having S-shaped chicanes, and thirdly in Si nanowires with engineered surfaces. The nanowire thermal conductance has been measured using longitudinal 3$\omega$ technique using a highly sensitive thermometry based on niobium nitride (NbN) \cite{bou2007,Heron1,Heron2,bou2006}.
The Si nanowires are fabricated from silicon on insulator (SOI) substrates by e-beam lithography. The total length of each structure has been purposely set to 10$\mu$m in order to easily compare the thermal measurement of all kinds of nanowires (see Fig.~\ref{fig1}). The section of the nanowires is 100~nm by 200~nm. The central symmetry of the structures is necessary for the 3$\omega$ method employed to measure the thermal conductance \cite{bou2007}. 

The temperature profile looks like a parabola with the hottest point located at the center of the nanowire. The heat generated at each points of the suspended structure flows from the center to the heat bath on both sides. The thermal conductance is generally measured using a power of a few tens of femtowatts dissipated in the niobium nitride transducer deposited on top of the nanowires. This power creates on average a temperature gradient smaller than 1~mK. The values of the measured thermal conductance are around a few universal value of thermal conductance $K_0=\pi^2k^{2}_{B}T/3h \sim 10^{-12} T$~W/K$^2$; a value used to normalized the data. 

The equation~\ref{Caseq} gives the correct order of magnitude for the thermal conductance since a $K$ close to $10^{-11}$~W/K is predicted at 3K and a value of 
$1.5\times 10^{-11}$W/K is measured on a straight nanowire. For the explanation of temperature variation of the thermal conductance it is needed to take into account the contribution of specular reflection by introducing the Ziman mean free path, and hence the parameter $p$. This has been described in details in the past \cite{Heron1,Heron2,her2010}. We would like to highlight here that the contribution of ballistic phonons to the thermal transport is significant. We have proposed a solution for thermal management that will not perturb diffusive transport but will affect only ballistic one. Specific suspended nanowires have been made including a double bend chicane structures of 400~nm long; the total extended length of the S-shaped nanowire itself being identical to the one of the straight nanowire (see Fig.~\ref{fig1}c and Fig.~\ref{fig2}). 

In recent works \cite{her2010,bla2014}, we have shown that the S-shaped nanowires have a much lower thermal conductance than straight one, as it can be seen in Fig.~\ref{fig2}. If the transport would be only diffusive, this constriction would have no effect at all on the thermal conductance. Actually, the severe reduction observed (reduction by more than 30 \%) demonstrates two significant facts: first, this illustrates the contribution of ballistic phonon to the thermal transport and second, including a chicane in a thermal conductor is a very efficient way of reducing the thermal transport. Phonons having long mean free path (bigger than the Casimir mean free path $\Lambda_{Cas}$) will be blocked and reflected by the chicane acting like a mirror for the long wave length acoustic plan waves. Inversely, short wavelength phonons that are more scattered by the nanowire surfaces won't be much affected by the chicane; their mean free path being already set by the section of the nanowire like in the Casimir regime. This concept has been successfully used to thermally isolate a suspended platform dedicated to sensitive thermal measurements \cite{kady}.

\subsection{Reduced thermal transport in corrugated nanowires}

\begin{figure}
\begin{center}
 \includegraphics[width=10cm]{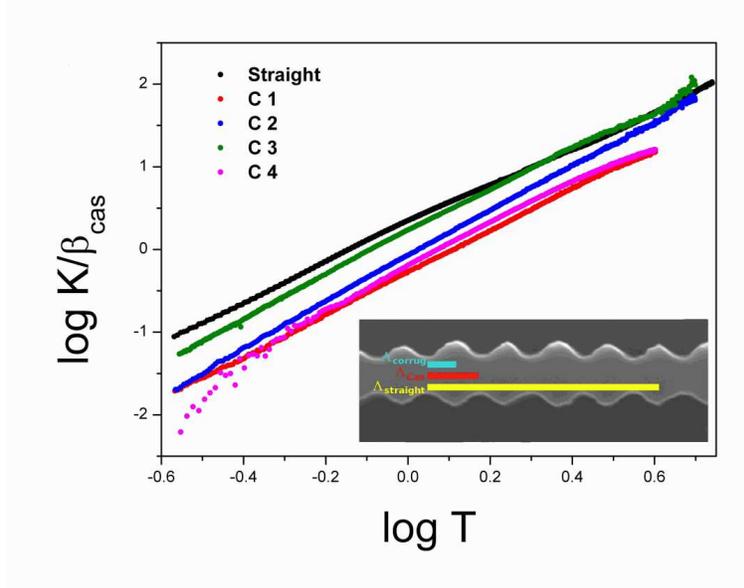}
 \end{center}
 \caption{Normalized thermal conductance versus temperature for a straight nanowire (Straight) and four corrugated nanowires (C1 to C4) in a log-log plot. The normalization coefficient $\beta_{Cas}$ corresponds to the Casimir limit value of the thermal conductance at 1~K. The corrugated nanowire C$_i$ can have much lower thermal conductance that smooth nanowire. The phonon MFP is reduced severely, even below the Casimir limit. As an illustration, in insert we compare the average phonon mean free path in the straight nanowire (in yellow), in the Casimir limit (in red), and as measured in corrugtaed nanowire (in blue). The corrugations reduce significantly the phonon mean free path.} 
 \label{fig3}
\end{figure}

In the previous section, we have seen how a macroscopic constriction can act as a phonon filter. In this section, we will illustrate how phonon backscattering or multiple scatterings can be introduced by using nano-engineered surfaces. In this respect, phononic crystals, defined as a regular modulation of the geometry of a phonon waveguide, have attracted significant attention in the past decade as a promising means to reduce phonon transport. In the present samples, periodic nanostructurations are used (see Fig.~\ref{fig3}) to induce phonon backscattering and reduce incoherently the thermal conductance; not coherently like in a real thermocrystal. We have reported that in that kind of engineered corrugated nanowires the phonon transport can be decrease even below the Casimir limit \cite{bla2014,bla2013}.

In order to compare thermal transport between various geometries of nanowires, the thermal conductances have to be normalized to a geometrical factor called the Casimir coefficient $\beta_{Cas}$, according to Eqs.~\ref{Caseq} and \ref{zimfp}. Fig.~\ref{fig3} reports the thermal conductance plotted as $K/\beta_{Cas}$. 

This normalization yields the temperature power law and allows a direct comparison between the thermal transport of each nanowire. Indeed, from Eqs.~\ref{Caseq} and \ref{zimfp}, the normalized conductance can be expressed as:
\begin{equation}
log (K/\beta_{Cas}) = a+b\hspace{0.1cm}log(T)
\label{kkcas1}
\end{equation}
where 
\begin{equation}
(K/\beta_{Cas})=\frac{1+p}{1-p} T^{b}
\label{kkcas2}
\end{equation}
From these two latter equations, we can extract the power law factor $b$ of the temperature variation of the thermal conductance, the value of the MFP $\Lambda_{eff}$ and the parameter $p$ \cite{bla2014,bla2013}.

The thermal conductance of the corrugated nanowires is clearly below the thermal conductance of straight nanowires even if the power law has not been change; being very close to 3 ($b\approx 2.6$). The presence of artificial periodic nanostructuration increases the scattering of phonon (backscattering, phonon trapping etc...). The surface roughness is not changed but the mean free path is clearly reduced. This can be depicted by ray-tracing Monte Carlo numerical simulations. Indeed, the multiple scattering of ballistic phonons can be seen through a corpuscular picture explaining the fact that the phonons are somehow trapped in the core of the nanowire. As a consequence, the phonon MFP that can be even smaller than the Casimir limit, i.e. smaller than the mean diameter of the nanowire. This is characterized by the necessity of putting a negative $p$ parameter to fit the data. Even if this special result does not involve phonon band engineering like in a real phononic crystal, it gives however a fantastic way of turning a single crystal semiconductor into a phonon glass.

\section{In the diffusive limit: nano-inclusions limits thermal conductivity in semiconductor}
\label{sec3}

After having shown in details the different ways of affecting the thermal transport in the presence of ballistic phonons, we will focus now on room temperature effects where the diffusive regime is governing the thermal conductivity.
We will show that engineering the phonon thermal properties is also possible with semiconductor at room temperature (RT). 
As compared to the ballistic limits, working at RT in the diffusive limit will have significant implications for real applications in managing heating, cooling or harvesting energy. 

The great challenges mostly rely on our capacity of improving the semiconductor materials performances. As on one hand a semiconductor is by far too thermally conductive (around 100~W/mK), on the other hand its Seebeck coefficient is generally quite high (above 200~$\mu$V/K). Then obtaining an efficient thermoelectric material having a high ZT based on a semiconductor is possible by decreasing significantly its thermal conductivity. 

The increase of the thermoelectric efficiency is feasible thanks to the difference of mean free path existing between phonons ($\sim$100~nm at 300~K) and electrons ($\sim$1~nm at 300~K). By introducing in a given material a structural disorder at the nanometer scale, it is possible to induce phonon diffusion without affecting the charge carriers justifying the innovative concept of \textit{electron crystal-phonon glass} material.

Indeed, in a semiconducting matrix, phonons contribute the most to the heat transport having an average mean free path of $\sim$100~nm at room temperature and a dominant wavelength of 1~nm at RT. With the objective of manipulating and controlling the phonon transport in a SC, the engineering of materials can then follow the two different approaches mentioned in the introduction: either by playing with phonon wavelength \cite{ila2014,nomura1,nomura2,wagner} or with the phonon mean free path \cite{bla2013,nomura2,wagner}. At room temperature, the most efficient effect will be the reduction of the phonon mean free path using a geometrical scattering approach. Indeed, as it has been recently experimentally demonstrated the phonon wavelength is far two small at RT for the nanostructuration to create sizable effects \cite{nomura2,wagner}.  

The best large scale applicability is expected from bulk 3D SC material that has been engineered at the low dimension for thermal management. One of the most important concepts is the \textit{electron crystal-phonon glass} being an inclusion-embedded semiconducting system, which shows great advantages and potentials \cite{kim}. For crystalline semiconductors, the ideal model would be a defect-free highly crystalline matrix containing nanoscaled inclusions with a wide diameter distribution. It can be theoretically predicted that as the inter-distance of the inclusions is comparable to the phonon mean free path in the matrix, the nano-inclusions can prohibit efficiently the phonon transport. Indeed, by introducing more phonon scattering processes we expect to get a germanium based \textit{electron crystal-phonon glass} material.

\begin{figure}
\begin{center}
 \includegraphics[width=6cm]{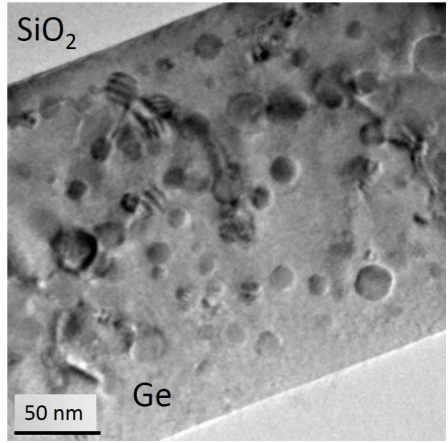}
 \end{center}
 \caption{TEM image of the germanium matrix with GeMn spherical nano-inclusions at 8~\% of Mn. The presence of numerous inclusions will significantly increase the phonon scattering without perturbing the electronic transport. The scale bar represents 50~nm. The distance between the nano-inlcusions is indeed smaller than the average phonon mean free path.} 
 \label{fig4}
\end{figure}

The materials of interest here are thin films of a single-crystalline germanium matrix embedded with randomly distributed Ge$_3$Mn$_5$ nano-inclusions grown by epitaxy on a germanium substrate (see Fig.~\ref{fig4}). The nano-inclusions are found to be nearly spherical with a diameter distribution of from 5 to 50~nm, depending on the growth parameters. Furthermore, manganese atoms act as p-type dopant in the Ge matrix, which ensures a good electrical conductivity of the material. 

The experimental measurements of the thermal conductivities of the Ge:Mn materials have been done using a highly sensitive differential 3$\omega$ technique \cite{liu2014}. Experimental results of 3$\omega$ measurements revealed strong reductions of the thermal conductivity, compared to the Ge bulk at room temperature, for Ge:Mn thin films containing different Mn concentrations.

The major experimental finding is a reduction of more than a factor of ten from the bulk value in the 8~\% of Mn in the Ge matrix. A thermal conductivity below 8~W/mK has been measured which has to be compared to the bulk thermal conductivity 65~W/mK. The reduction is fully ascribed to the presence of evenly distributed phonon scatterers than significantly reduce the phonon mean free path. 

\section{Conclusions}
\label{sec4}

Different concepts of phonon thermal transport reduction have been evidenced experimentally from the ballistic limit to the diffusive limit. These new paradigms will be fruitfully applied for the development of new materials or devices in thermoelectricity and energy harvesting in general. It seems that now, we are facing a kind of intrinsic limit in that reduction by non-coherent effects, a lower limit close to the alloy value of thermal conductivity. Unfortunately, engineering the phonon band by opening gaps like in a phononic crystal seems not perfectly appropriate for room temperature thermal management, phonon wavelength being far too small as compared to the length accessible by nowadays fabrication means. 

Other technological solutions may be found in playing also on the electrical properties of semiconductor based systems to improve the Seebeck coefficient. Indeed as proposed by Hicks and Dresselhaus \cite{dress1} it could be possible to increase the Seebeck coefficient by playing on the electronic density of state at the Fermi level. Then, engineering also the electronic properties by various means (nanostructuration, adding quantum dots, etc...) seems absolutely needed if one wants to boost the performances of future thermoelectric materials.

\section*{Acknowledgments}
We acknowledge technical supports from Nanofab, the Cryogenic and the Electronic facilities and the Pole Capteur Thermom\'etrique et Calorim\'etrie of Institut N\'eel for these experiments. Funding for this project was provided by a grant from La R\'egion Rh\^one-Alpes (Cible), by the Agence Nationale de la Recherche (ANR) through the project QNM no. 0404 01, by the European projects: MicroKelvin FP7 low temperature infrastructure Grant no. 228464 and MERGING Grant no. 309150.

\end{document}